\documentstyle[12pt,cite,epsf,axodraw]{article}
\title{Neutrino Masses from Non-minimal
Gravitational Interactions of Massive Neutral Fermions}

\author{Abhinav Gupta\thanks{E--mail : abh@ducos.ernet.in}, 
Namit Mahajan\thanks{E--mail : nm@ducos.ernet.in, 
nmahajan@physics.du.ac.in} and
Amitabha Mukherjee\thanks{E--mail : am@ducos.ernet.in}\\
	{\em Department of Physics and Astrophysics,} \\
	 {\em University of Delhi, Delhi-110 007, India.}}

\begin{document}
%\doublespacing
\maketitle
%\large
%\newcommand{\del}{\mbox} 
\begin{abstract}
A new mechanism is proposed for generating neutrino masses radiatively 
through a non-minimal coupling to gravity of fermionic bilinears 
involving massive neutral fermions.  Such coupling terms can arise in theories
where the gravity sector is augmented by a scalar field. They necessarily 
violate the principle of equivalence, but such violations are not ruled
out by present experiments.  It is shown that the proposed 
mechanism is realised most convincingly in theories of the Randall-
Sundrum type, where gravity couples strongly in the TeV range.
The mechanism has the potential for solving both the solar
and atmospheric neutrino problems. The smallness of neutrino masses in
this scenario is due to  the fact that the interaction of the 
massive neutral fermions arises entirely from higher-dimensional operators
in the effective Lagrangian.
\\ \\
{\bf Keywords}: Neutrino Masses, Non-minimal coupling, 
massive neutral fermion \\ \\
{\bf PACS}: 14.60.Pq, 14.60.St     
\end{abstract}
%\vskip 0.5cm
\begin{section}*{Introduction}
It is well known that non-zero neutrino masses can result in
neutrino oscillations.
There seems to be strong experimental evidence for such 
oscillations \cite{exp}. Several models have been proposed to 
generate the neutrino mass matrix \cite{models}. Most of these models
modify the particle content of the Standard Model (SM) and introduce 
new fields and/or new interactions. In these
scenarios gravity couples universally to matter, preserving
the Equivalence Principle (EP). \\ \\ 
\indent The EP, when extrapolated to microscopic scales, is a theoretical
prejudice rather than an empirical principle. Earlier experimental
 tests of EP
\cite{EP, long} placed fairly stringent bounds on 
the strength of a possible
EP-violating term at distance scales larger than a metre.
The latest measurements have further pushed the lower limit of the
range to about a millimetre, and, as summarised in \cite{long},
`` very little is known about gravity at length scales below a few 
millimetres``. Thus substantial violations of the EP remain possible
at sub-atomic scales. 
In particular, EP violation is generic to many 
higher dimensional scenarios \cite{nussinov}, and the possibility of
 generating neutrino 
masses in such theories has been considered \cite{nima-moha}. 
 The present study is concerned 
with the possiblity of generating realistic neutrino masses through
violation of the equivalence principle (VEP) at length scales much below
those probed by present experiments \cite{EP} . \\ \\
\indent  The possibility of VEP 
as a source of neutrino oscillations has been explored in
the literature \cite{gasperini}- \cite{dv1}.
 Most of these studies \cite{gasperini} assume non-universal
coupling of gravity to different neutrino flavours, leading to possible 
oscillations in massless/massive neutrinos through their asymmetrical 
interaction with the macroscopic gravitational potential.
However, Adunas et.al \cite{dv1} argue that possible quantum limitations
on EP follow from a fundamental uncertainity in measurement of mass.
\footnote{Gravitational effects in neutrino oscillations have
also been studied from a different perspective, viz. that of 
gravitationally induced quantum phases \cite{dv2}.} \\ \\
\indent We explore the possibility of generating
neutrino masses through a violation 
of EP by introducing non-minimal gravitational interaction terms 
in the action involving
neutrinos and Massive Neutral Fermions (MNFs). 
We show that it is possible to generate neutrino
masses by the addition of such a non-minimal term in theories where the
gravity sector is enlarged by additional massive scalar field(s). 
The TeV scale
 quantum gravity theories \cite{nima-rs} are found to be the natural 
candidates for realising this mechanism.   
\end{section}
\begin{section}*{The Model}
Consider a theory of gravity with a scalar field that couples to the 
matter fields through the trace of the energy-momentum tensor. The total 
action involving gravitation and matter coupling minimally is 
\begin{equation}
S_{total} = S_{gravity} + \int d^4x~\sqrt{-g}~{\cal{L}}_{matter} 
\end{equation}
The matter Lagrangian is constructed so that, in addition to the usual
 SM fields, there is a MNF ($S_1$) with Dirac 
mass ($m_{S_1}$), as well as
kinetic terms for the right handed partners ($\nu^i_R$) of 
the SM neutrinos with the index $i$ running over neutrino flavours
 $e$, $\mu$, $\tau$.
The MNF and the right handed neutrinos are $SU(2)_L\times
U(1)_Y$ singlets thus having no SM interactions. The non-SM
interactions are in the gravity sector alone, induced by a non-minimal
coupling leading to an additional term in the action of the form  
\begin{equation}
S_{NM} = -\frac{1}{\Lambda^2}\sum_if^i\int d^4x\sqrt{-g}R
\Bigg[~{\bar{\psi^i}}_L\tilde{\Phi}{S_1}_R 
 - c^i\frac{v}{\sqrt{2}}{\bar{S_1}_L}\nu^i_R
 + h.c~\Bigg]
\end{equation}
where $\psi_L^i$ is the $i^{th}$ SM lepton doublet, $\tilde{\Phi} = 
\iota\Phi^{\ast}$ is the conjugate Higgs doublet, $v$ is the 
Higgs vacuum expectation value (VEV) and $R$ is the Ricci
scalar. The coefficients $f^i$ and $c^i$ are arbitrary dimensionless
parameters determining the strengths with which different
flavours couple in the above non-minimal fashion.
 $\Lambda$ is the characteristic mass-scale of the theory of 
gravity under consideration, determined by some higher theory which 
becomes relevant above this scale. This term is invariant under 
$SU(2)_L\times U(1)_Y$ transformations and  violates EP. The extent of 
VEP by different neutrino flavours is characterised by $f^i$ and $c^i$.
Such higher order non-minimal terms may be generated as a result
 of quantum effects of the higher theory of which this theory is a 
low-energy manifestation {\footnote{In general, in such theories
other terms with mass dimension $>$ 4 may also be induced.
Such terms have been extensively studied and exploited to 
generate neutrino masses \cite{akhmedov}. We 
proceed on the assumption that, for some choice of parameters in the
 underlying theory, the other terms are negligible.}}.
 This additional term leads to a modification 
of the energy-momentum tensor 
\begin{equation}
\Delta~T_{\mu\nu} = \frac{-2}{\Lambda^2}\sum_if^i
(\eta_{\mu\nu}\partial_{\alpha}\partial^{\alpha} -  
\partial_{\mu}\partial_{\nu}) 
\Bigg[~{\bar{\psi^i}}_L\tilde{\Phi}{S}_R 
 - c^i\frac{v}{\sqrt{2}}{\bar{S}_L}\nu^i_R
 + h.c.~\Bigg]
\end{equation}
Assuming the coupling of the scalar field ($\phi$ with mass 
$M_{\phi}$) to the trace of 
energy-momentum tensor is of the form
\begin{equation}
{\cal{L}}_{int} = -\frac{\phi}{\Lambda}(T_{matter})_{\mu}^{\mu}
\end{equation}
the change in $T_{\mu\nu}$ results, after SM spontaneous symmetry
 breaking (SSB), to an interaction of the scalar to neutrinos
of the form
\begin{equation}
{\cal{L}}_{int} = \sum_i\frac{6f^iM_{\phi}^2v}{\sqrt{2}\Lambda^3}
\Bigg[~{\bar{\nu^i}}_L{S_1}_R 
 - c^i{\bar{S_1}_L}\nu^i_R
 + h.c~\Bigg]~\phi
 \end{equation}
For reasons evident later, we choose the parameters $c^i$ to be 
all equal to some constant $c$.
 From the interaction Lagrangian it is evident that 
a neutrino mass matrix would be induced by the radiative 
process shown in Figure 1.   
\begin{center}
 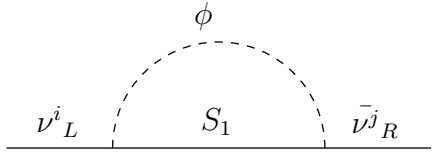
\begin{figure}[htb]
\vspace*{-10ex}
\hspace*{2em}
\begin{tabbing}
\begin{picture}(155,120)(-5.0,-20)
\Line(150,0)(310,0)
	\Text(170,10)[c]{${\nu^i}_L$}
	\Text(230,10)[c]{$S_1$}
	\Text(290,10)[c]{$\bar{{\nu^j}}_R$}
\DashCArc(230,0)(40,0,180){3}
	\Text(225,50)[c]{$\phi$}

\end{picture}
\end{tabbing}
\caption{Typical Feynman diagram (evaluated at zero external momenta)
contributing to the $ij$ entry of the mass matrix.}
\end{figure}
\end{center}
The mass matrix for three generations has the following structure
\begin{equation}
{\cal{M}}_{\nu} = \delta~ \left( \begin{array}{ccc}
{f^e}^2&f^ef^{\mu}&f^ef^{\tau}\\
f^ef^{\mu}&{f^{\mu}}^2&f^{\mu}f^{\tau}\\
f^ef^{\tau}&f^{\mu}f^{\tau}&{f^{\tau}}^2
\end{array} \right) 
\end{equation}
where 
\[
\delta = \frac{c~m_{S}}{16\pi^2}\left(\frac{6~M_{\phi}^2~v}
{\sqrt{2}~\Lambda^3}\right)^2\left(\frac{1}{M_{\phi}^2-m_S^2}\right)
\Bigg[M_{\phi}^2\log\left(\frac{\Lambda^2}{M_{\phi}^2}\right)
- m_S^2\log\left(\frac{\Lambda^2}{m_S^2}\right)\Bigg]
\]
This matrix has two zero eigenvalues and one non-zero eigenvalue $m_3$,
given by
\begin{equation}
m_3 = \delta({f^e}^2 + {f^{\mu}}^2 + {f^{\tau}}^2)   
\end{equation}
The weak interaction gauge eigenstates $\nu^i$ are related to the
mass eigenstates $\nu^a$ ($a = 1, 2, 3$) by the mixing
matrix $U$ as 
\[
\nu^i = \sum_a U^{ia}\nu^a
\]
where the mixing matrix is
\begin{equation}
U = \left( \begin{array}{ccc}
s_{\alpha}c_{\theta}+c_{\alpha}s_{\theta}\gamma
&-s_{\alpha}s_{\theta}+c_{\alpha}c_{\theta}\gamma
&\rho c_{\alpha}\gamma \\
-c_{\alpha}c_{\theta}+s_{\alpha}s_{\theta}\gamma
&c_{\alpha}s_{\theta}+s_{\alpha}c_{\theta}\gamma
&\rho s_{\alpha}\gamma\\
-\rho s_{\theta}\gamma
&-\rho c_{\theta}\gamma
&\gamma
\end{array} \right)
\end{equation}
where $c$ and $s$ with subscript are the cosine and sine of 
the subscript respectively, 
$\rho = {({f^e}^2+{f^{\mu}}^2)^{\frac{1}{2}}}/{f^{\tau}}$, 
$\tan\alpha = f^{\mu}/f^e$ and 
$\gamma = 1/(1+\rho^2)^{\frac{1}{2}}$. The angle $\theta$ corresponds
to possible rotations in the degenerate $\nu^1$-$\nu^2$ sub-space.\\ \\
\indent The above analysis assumes the same value $c$ for the parameters
$c^i$ as mentioned before. But it is easy to show that a different value for 
each $c^i$ does not alter the mass spectrum as such, since one still
gets two zero eigenvalues; the non-zero eigenvalue, 
though, is altered. In this minimal framework, the atmospheric
neutrino problem is solved by $\nu^{\mu} \rightarrow \nu^{\tau}$ 
oscillations, with the assumption that $f^e << f^{\mu}, f^{\tau}$
implying that $\nu^e-\nu^{\mu}$ and $\nu^e-\nu^{\tau}$ mixing is 
negligible compared to $\nu^{\mu}-\nu^{\tau}$ mixing. \\ \\
%%%%%%%%%%%%%%%%%%%%%%%%%%%%%%%%%%%%%%%%%%%%%%%%%%%%%%%%%%%%%%%%%%%%
\indent To solve both the atmospheric and solar neutrino problems
in a three flavour framework, we need two mass scales corresponding
to the different length  scales involved in the two problems.
Therefore, the degeneracy of the $\nu^1$-$\nu^2$ sub-space has to be 
lifted. This can be achieved by extending this minimal framework 
by including 
two more MNFs ($S_2$ and $S_3$) having a mass
similar to $m_{S_1}$. We thus have three MNFs
$S_m$ ($m = 1, 2, 3$) interacting non-minimally to generate the 
neutrino mass matrix.
The Lagrangian for the extended system is
\begin{equation}
{\cal{L}}_{int} = \frac{6M_{\phi}^2v}{\sqrt{2}\Lambda^3}
    \sum_{m=1}^3 \sum_{i=e,\mu,\tau}
\Bigg[~f_m^i({\bar{\nu^i}}_L{S_m}_R 
  - c_m^i{\bar{S_m}}_L\nu^i_R) + 
    h.c.~\Bigg]~\phi
 \end{equation}
For simplicity we choose $c_m^i = 1$ and the masses for $S_m$ 
to be the same. \\

\indent The following choice of parameters
\begin{eqnarray}
f_1^{\tau} = - f_1^{\mu} = \frac{1}{\sqrt{2}}f_1^e \\ \nonumber
f_2^{\tau} = - f_2^{\mu} = -\frac{1}{\sqrt{2}}f_2^e \\ \nonumber
f_3^{\tau} = f_3^{\mu} \hskip 0.3cm and \hskip 0.3cm f_3^e = 0
\end{eqnarray} 
reduces the mass matrix to the form
\begin{equation}
{\cal{M}}_{\nu} = \delta~ \left( \begin{array}{ccc}
{f_1^e}^2+{f_2^e}^2&\frac{1}{\sqrt{2}}({f_2^e}^2-{f_1^e}^2)
&\frac{1}{\sqrt{2}}({f_1^e}^2-{f_2^e}^2)\\
\frac{1}{\sqrt{2}}({f_2^e}^2-{f_1^e}^2)&
\frac{1}{2}({f_1^e}^2+{f_2^e}^2) +{f_3^{\mu}}^2&
-\frac{1}{2}({f_1^e}^2+{f_2^e}^2) +{f_3^{\mu}}^2\\
\frac{1}{\sqrt{2}}({f_1^e}^2-{f_2^e}^2)&
-\frac{1}{2}({f_1^e}^2+{f_2^e}^2) +{f_3^{\mu}}^2&
\frac{1}{2}({f_1^e}^2+{f_2^e}^2) +{f_3^{\mu}}^2
\end{array} \right) 
\end{equation}
with eigenvalues
\begin{equation}
m_1 = 2\delta {f_1^e}^2 \hskip 0.5cm m_2 = 2\delta {f_2^e}^2
\hskip 0.5cm m_3 = 2\delta {f_3^{\mu}}^2
\end{equation}
The mixing matrix for this choice of parameters is
\begin{equation}
U = \left( \begin{array}{ccc}
\frac{1}{\sqrt{2}}&\frac{1}{\sqrt{2}}&0\\
-\frac{1}{2}&\frac{1}{2}&\frac{1}{\sqrt{2}}\\
\frac{1}{2}&-\frac{1}{2}&\frac{1}{\sqrt{2}}
\end{array} \right)
\end{equation}
This mixing matrix, which is a special case of a general $3\times 3$
mixing matrix, is referred to as the {\it 'bimaximal mixing'} solution
  \cite{bimax}
 to the solar and atmospheric neutrino problems. The observed zenith-angle
independence of e-like events at Super-Kamiokande places strong
constraints on the mixing matrix. The essentially unique solution to 
these constraints \cite{dv4} has one zero element and depends on a 
single angle. The bimaximal solution corresponds to choosing this angle
to be equal to $\pi/4$.
 In general,
the parameters of the theory  can be altered to accommodate other 
possible solutions
in a three flavour framework.\\ \\ 
\indent The above solution is consistent with
the following constraints \cite{bimax} 
on the mass eigenvalues with the hierarchy
$m_1 < m_2 < m_3$ 
\begin{eqnarray}
\Delta_{atm.} &=& m_3^2 - m_2^2 \sim 3.5 \times 10^{-3} ~eV^2 \\ \nonumber
\Delta_{sol.} &=& m_2^2 - m_1^2 \sim 10^{-5} ~eV^2 ~(Large ~Angle ~MSW
 ~solution) 
\end{eqnarray}
for the atmospheric and solar neutrino problems respectively. This puts 
constraints on the masses  of MNFs and the 
parameters $f_m^i$ which are a measure of the extent of VEP. A large 
value of $m_S$ would result in weaker violation of EP to be 
consistent with equation (14).\\ \\

\indent The formalism assumes the existence of a massive scalar in the 
gravity sector coupling to the trace of the matter energy-momentum tensor.
This requirement is fulfilled in the current TeV scale gravity theories 
\cite{nima-rs}. In the Randall-Sundrum (RS) scenario, a massive scalar, 
the radion,
couples to the matter fields in the desired fashion making it a possible
candidate to realise this kind of mass generation for neutrinos.
In this scenario, $\Lambda \sim 10$ TeV and $M_{\phi} \sim 250$ GeV are 
typical scales, leading to $\delta \sim m_S \times 10^{-10}$. Equation (14) 
then leads to
\begin{eqnarray}
{f_2^e}^4 - {f_1^e}^4 &\sim& 10^{-8} \\ \nonumber
{f_3^{\mu}}^4 - {f_2^e}^4 &\sim& 10^{-6} \hskip 1cm for ~~m_S \sim 100 ~GeV
\end{eqnarray}     
and 
\begin{eqnarray}
{f_2^e}^4 - {f_1^e}^4 &\sim& 10^{-10} \\ \nonumber
{f_3^{\mu}}^4 - {f_2^e}^4 &\sim& 10^{-8} \hskip 1cm for ~~m_S \sim 1~TeV
\end{eqnarray}
It is evident that a larger mass for MNFs leads to
a weaker VEP. \\
\indent The $f's$ can be further constrained by using the neutrino
mass bounds as given in \cite{PDG}. These bounds imply the
following constraints on $f's$ :
\begin{equation}
{f_1^e}^2 < 7.5\times 10^{-10} \hskip 1cm 
{f_2^e}^2 < 8\times 10^{-5} \hskip 1cm  
{f_3^{\mu}}^2 < 9\times 10^{-4} 
\end{equation}
for $m_S = 100$ GeV.
It can be seen that these values are in agreement with Equation (15).
Similar bounds for $f's$ can be obtained for other values of $m_S$.
\end{section}
\begin{section}*{Discussion}
\indent The above analysis has been carried out in a three neutrino
flavour framework with bimaximal mixing parameters. If LSND results
are confirmed in future, it would imply the possible existence of a
 light sterile
neutrino. Such a sterile neutrino would interact with the MNFs in a manner
similar to ordinary neutrinos, acquiring mass and generating a third 
mass scale accounting for the LSND results (if confirmed).\\ \\
\indent In this formalism, since the mass matrix is generated only through 
non-standard gravitational interactions
of the massive neutral fermions and ordinary neutrinos, 
the SM precision tests are not 
disturbed below the characteristic scale, $\Lambda$, of the effective theory.
 The constraints on the parameters have been
obtained for the RS scenario, but the formalism applies to a generic
class of scalar-tensor theories of gravitation in which the scalar 
couples to the trace of the energy-momentum tensor of matter.\\ \\
\indent It is evident that such a mass generation is not possible if the 
scalar field is massless. The non-minimal term would also give rise to
interactions of neutrinos and MNFs with ordinary massless gravitons.
But to lowest order in the effective coupling, the masslessness of the 
graviton forbids
such an interaction. The non-zero mass of the scalar field ensures that
the non-minimal interaction does not lead to any long-distance corrections
to ordinary gravitational interactions. Thus, the constraints obtained 
above are not in conflict with millimetre scale tests of EP. \\ \\
\indent A question which may be asked in such models relates to 
proton decay. The explicit lepton number violation considered here,
it can be argued, will lead to baryon number violation and hence to
an unacceptably rapid proton decay. However, it has been shown 
in the context of MSSM by Dines {\it {et.al}}
\cite{nima-moha}, that there is a higher dimensional mechanism
involving Kaluza-Klein selection rules which enables us to cancel
the proton decay diagrams to all orders in perturbation theory.
The authors also argue that within the context of string theory, 
it is possible to circumvent the proton decay problem through other 
model-dependent mechanisms. We hope that one can, in principle,
find such a way to circumvent the problem in the context of the 
class of models described in this paper.
\end{section}
\begin{section}*{Acknowledgements}
We would like to thank Anjan Joshipura
for helpful discussions.
A.~G would like to thank C.S.I.R, India while N.~M thanks U.G.C, India
for fellowship. 
\end{section} 

\end{document}